\documentclass[pre,aps,amssymb,amsmath,showpacs,superscriptaddress,twocolumn]{revtex4}

\usepackage{amsmath}
\usepackage{amssymb}
\usepackage{epsfig}
\usepackage{amscd}
\usepackage{graphicx,psfrag,xspace}
\usepackage{float}
\usepackage[normalem]{ulem}

\usepackage{bm}
\usepackage{bbm}

\begin{document}

\title{Clustering of branching Brownian motions in confined geometries}
\author{A.~Zoia}
\email{andrea.zoia@cea.fr}
\affiliation{CEA/Saclay, DEN/DANS/DM2S/SERMA/LTSD, 91191 Gif-sur-Yvette, France}
\author{E.~Dumonteil}
\affiliation{CEA/Saclay, DEN/DANS/DM2S/SERMA/LTSD, 91191 Gif-sur-Yvette, France}
\author{A.~Mazzolo}
\affiliation{CEA/Saclay, DEN/DANS/DM2S/SERMA/LTSD, 91191 Gif-sur-Yvette, France}
\author{C.~de Mulatier}
\affiliation{CEA/Saclay, DEN/DANS/DM2S/SERMA/LTSD, 91191 Gif-sur-Yvette, France}
\affiliation{CNRS - Universit\'e Paris-Sud, LPTMS, UMR8626, 91405 Orsay Cedex, France}
\author{A.~Rosso}
\affiliation{CNRS - Universit\'e Paris-Sud, LPTMS, UMR8626, 91405 Orsay Cedex, France}

\begin{abstract}
We study the evolution of a collection of individuals subject to Brownian diffusion, reproduction and disappearance. In particular, we focus on the case where the individuals are initially prepared at equilibrium within a confined geometry. Such systems are widespread in physics and biology and apply for instance to the study of neutron populations in nuclear reactors and the dynamics of bacterial colonies, only to name a few. The fluctuations affecting the number of individuals in space and time may lead to a strong patchiness, with particles clustered together. We show that the analysis of this peculiar behaviour can be rather easily carried out by resorting to a backward formalism based on the Green's function, which allows the key physical observables, namely, the particle concentration and the pair correlation function, to be explicitly derived.
\end{abstract}


\maketitle

\section{Introduction}

Many relevant systems occurring in physics and in biology can be described in terms of a collection of individuals undergoing branching random walks, where stochastic spatial displacements are coupled to some reproduction-disappearance mechanism~\cite{legall, athreya, harris}. The evolution of the neutron population in a nuclear reactor in the presence of multiplication due to fission events provides a relevant example~\cite{pazsit, williams, osborn, bell_nuc, pal, zoia1, zoia2}. In the context of life sciences, models of diffusion with birth-death events of the Galton-Watson type~\cite{zhang, meyer} (the so-called `Brownian bugs') have been successfully applied to, among others, the dynamics of bacterial colonies~\cite{jagers, golding, houchmandzadeh_pre_2002, houchmandzadeh_prl, houchmandzadeh_pre_2009}, the spread of epidemics~\cite{bailey, pnas}, the mutation-propagation of genes~\cite{lawson, bertoin, sawyer, cox, dawson}, and the spatial patterns of ecological communities~\cite{young, murray, spatial_eco}. Generally speaking, individuals may interact with each other~\cite{flierl, lopez}, which would make their evolution intrinsically non-linear. For the sake of simplicity, we will focus on neutral populations, whose individuals interact with the host medium but not with each other. This assumption is surely legitimate for fairly diluted systems, such as neutrons in nuclear reactors, whose number density is much smaller than that of the surrounding nuclei~\cite{pazsit, williams}. In the context of ecology, neutral evolution has been evoked in order to separately investigate the effects of birth, death and migration without having to explicitly take into account the influence of environmental parameters (spatial heterogeneities) and individual interactions (social behaviour)~\cite{young, houchmandzadeh_prl, houchmandzadeh_pre_2009}. For random walk models of epidemics, neutrality would require the nonlinear effects due to the depletion of the susceptibles to be neglected, which is a common assumption during the outbreak phases~\cite{jagers, pnas}. Even under the simplifying hypothesis of neutral evolution, deriving precise asymptotic estimates for branching processes often demands a great amount of ingenuity~\cite{derrida, derrida_barrier, berestycki_one, berestycki_two, berestycki_three}.

\begin{figure}[t]
\centerline{\epsfclipon \epsfxsize=9.0cm
\epsfbox{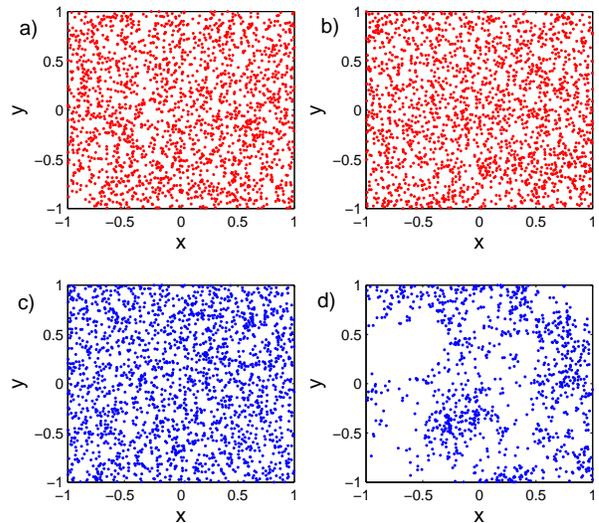} }
\caption{(Color online). Monte Carlo simulation of a collection of $N=2\cdot 10^3$ individuals in a two-dimensional box of half-size $L=1$ with reflecting boundary conditions. Each particle undergoes a regular Brownian motion with diffusion coefficient $D=10^{-2}$. At exponentially distributed times, with rate $\lambda = 1$, the walkers disappear and give rise to a random number $k$ of identical and independent descendants (a Galton-Watson birth-death mechanism), with probability $q_k$, each behaving as the parent particle. For Figs.~$a)$ and $b)$ (red) we have chosen $q_1=1$, which means that particles can only diffuse, whereas for Figs.~$c)$ and $d)$ (blue) we have chosen $q_0=q_2=1/2$, which means that particles can branch (giving rise to two descendants) or be absorbed with equal probability (the so-called binary branching Brownian motion~\cite{cox}). In either case, the average number of descendants per reproduction event is equal to one. Figs.~$a)$ and $c)$ illustrate the initial configurations at time $t=0$ for the two simulations, respectively: in both cases, the $N$ particles are uniformly distributed in space over the box. The behaviour of the two systems at time $t=100$ is displayed in Figs.~$b)$ and $d)$, respectively. In the system with purely diffusive Brownian motions, at a later time the particle positions are just shuffled with respect to the initial configuration because of the random motions of the walkers. Fluctuations are Poissonian and do not sensibly affect the particle distribution, which stays uniform over the box. The evolution of the system with branching Brownian motions is considerably different: at a later time, the walker positions display a strong patchiness, with individuals clustered together. In this latter case, fluctuations are non-Poissonian and deeply affect the behaviour of the particle distribution.}
\label{fig1}
\end{figure}

Because of the combined effect of the spatial displacements and of the reproduction-disappearance mechanism, the local number $n_{V_i}(t)$ of individuals in the system at a given site ${V_i}$ in the viable phase space at time $t$ is subject to fluctuations around the average value, and so is the total number of individuals. In a deterministic approach, knowledge of $\mathbb{E}_t[ n_{V_i}]$ (i.e., the ensemble-averaged number of individuals) is assumed to be sufficient so as to characterize the system evolution~\cite{williams, young, houchmandzadeh_prl, houchmandzadeh_pre_2009}. This stems from assuming that fluctuations affecting the population are Poissonian, and become negligible when $\mathbb{E}_t[ n_{V_i}]$ is sufficiently large. In sharp contrast with this prediction, the spatial distribution of such individuals has been shown to possibly display a strong `patchiness', with walkers clustered together~\cite{zhang, meyer, young, houchmandzadeh_prl, houchmandzadeh_pre_2002, houchmandzadeh_pre_2009}. A numerical example obtained by Monte Carlo simulation is illustrated in Fig.~\ref{fig1}. The pioneering theoretical work performed in the context of mathematical ecology has revealed that the hypothesis of Poissonian fluctuations actually fails in the presence of branching: spatial correlations induced by the parent-child coupling become relevant whenever diffusion is not sufficient to smooth out such inhomogeneities~\cite{cox, zhang, meyer}. These phenomena are enhanced in particular in low dimensional systems ($d \le 2$)~\cite{cox, houchmandzadeh_pre_2002, houchmandzadeh_pre_2009}. Neutral clustering phenomena have been reported to occur in laboratory experiments and numerical simulations involving ecological communities~\cite{young, houchmandzadeh_prl, houchmandzadeh_pre_2002, houchmandzadeh_pre_2009}. In the context of reactor physics, though clustering of neutron populations has never been explicitly considered so far~\cite{dumonteil_ane}, the role of correlation-induced fluctuations has been extensively investigated in nuclear systems operated at very low power~\cite{williams, osborn, pazsit, bell, natelson_theory, natelson_experiment}. In all such cases, it has been shown that a deterministic approach to the description of the population behaviour would be meaningless, since the fluctuations of the local particle number may attain the same order of magnitude as the average particle number itself~\cite{williams, natelson_theory, osborn, young, houchmandzadeh_pre_2009}.

So far, mathematical modelling of clustering has mostly focused on the case of very large populations diffusing on unbounded domains, the so called thermodynamic regime~\cite{cox, zhang, meyer, young, houchmandzadeh_prl, houchmandzadeh_pre_2002, houchmandzadeh_pre_2009}. In many practical applications, however, one is interested in studying populations composed of a finite number of individuals and evolving in confined geometries. Neutrons in a nuclear reactor, for instance, are confined for safety reasons by reflecting and absorbing barriers that prevent radiation from escaping~\cite{williams, pazsit, bell}. In this work we show that the analysis of the fluctuations of branching Brownian motions in confined geometries (with arbitrary boundary conditions) can be rather easily carried out based on a backward formalism. In particular, the physical observables of interest, namely, the particle concentration and the pair correlation function, can be obtained in terms of the Green's function related to the underlying stochastic process. In Sec.~\ref{physical} we will derive the expressions for the physical observables, which will be then used in Sec.~\ref{fluctuations_around_equilibrium} so as to investigate the fluctuations around equilibrium for a collection of $N$ such individuals. In Sec.~\ref{one_dimensional} we will illustrate the proposed formalism on a simple example involving branching processes in a one-dimensional box. Conclusions are finally drawn in Sec.~\ref{conclusions}. Technical details and calculations are left to a series of Appendices.

\section{The physical observables}
\label{physical}

Consider a collection ${\cal S}$ of $N$ particles initially located at random positions ${\mathbf x}^1_0$, ${\mathbf x}^2_0$, ${\mathbf x}^3_0$, $\cdots$, ${\mathbf x}^N_0$ at time $t_0=0$, with associated density $P({\mathbf x}^1_0, {\mathbf x}^2_0, \cdots , {\mathbf x}^N_0)$. If the starting points are independently and identically distributed, we can factorize $P({\mathbf x}^1_0, {\mathbf x}^2_0, \cdots , {\mathbf x}^N_0) = \prod_{k=1}^N p({\mathbf x}^k_0)$. The walkers undergo random displacements (independently of each other) and are subject to random reproduction-disappearance events. In order to fix the ideas, in the following we will assume that the random displacements can be approximated by a regular $d$-dimensional Brownian motion  with diffusion coefficient $D$. This is a reasonable hypothesis for living organisms (as far as the support is sufficiently homogeneous)~\cite{jagers, weiss}, and holds also for neutrons in the so-called diffusion regime (in the absence of localized sources or sinks, and when scattering dominates over absorption)~\cite{williams, pazsit}. We will furthermore assume that at exponentially distributed times, with rate $\lambda$, each walker undergoes a Galton-Watson reproduction event: the particle disappears and is replaced by a random number $k$ of identical and independent descendants, distributed according to the probability $q_k$ and behaving as the parent particle (disappearance is taken into account by the event $k=0$). Such kind of stochastic process defines a branching Brownian motion~\cite{legall, athreya, harris}.

\begin{figure}[t]
\centerline{\epsfclipon \epsfxsize=9.0cm
\epsfbox{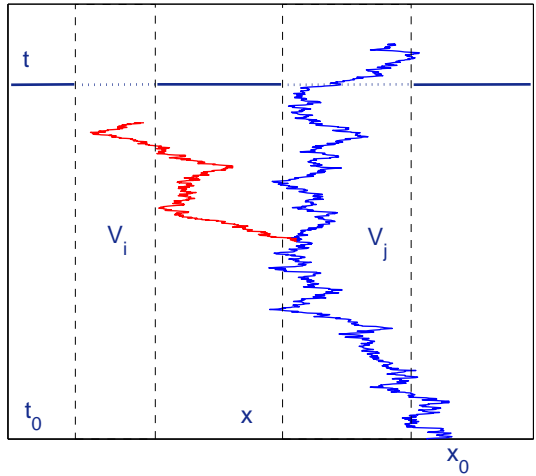} }
\caption{(Color online). Example of realization of a branching Brownian motion in one dimension. A single walker starts to diffuse from position $x_0$ at time $t_0=0$. At a later time, a branching event occurs, and a new independent Brownian motion starts to diffuse. At the observation time $t$, one of the two walkers is found in the region $V_j$, whereas the other has been absorbed at an earlier time and does not contribute to the counting process.}
\label{fig2}
\end{figure}

The individuals evolve in a $d$-dimensional domain $V$ with given boundary conditions on $\partial V$. We would like to characterize the statistical behaviour of the random number of particles $n_{V_i}$ that are found in a volume $V_i \subseteq V$ of the viable space at a given time $t$. Actually, in view of assessing the correlations of our system, we are more generally interested in determining the simultaneous detection at two volumes $V_i \subseteq V$ and $V_j \subseteq V$ at time $t$ (see Fig.~\ref{fig2}). The relevant physical observables are thus the average particle number at a given detector located at $V_i$, namely, $\mathbb{E}_t[ n_{V_i} |{\cal S}]$, and the correlations between two detectors respectively located at $V_i$ and $V_j$, namely, $\mathbb{E}_t[ n_{V_i} n_{V_j} |{\cal S}] $, when the process is observed at a time $t>t_0$.

The local particle concentration $c$ at a site ${\mathbf x}_i$ is then defined by centering the volume $V_i$ at ${\mathbf x}_i$ and taking the volume size $V_i \to 0$, namely,
\begin{equation}
c_t({\mathbf x}_i) = \lim_{V_i \to 0 } \frac{\mathbb{E}_t[ n_{V_i} |{\cal S}]}{V_i}.
\label{def_concentration_c}
\end{equation}
The quantity $c_t({\mathbf x}_i) d {\mathbf x}_i $ represents by definition the average number of particles to be found in a small volume $d {\mathbf x}_i $ around position ${\mathbf x}_i$ at time $t$. The local correlations $h$ between a site ${\mathbf x}_i$ and a site ${\mathbf x}_j$ are similarly defined by centering the volume $V_i$ at ${\mathbf x}_i$ and the volume $V_j$ at ${\mathbf x}_j$, respectively, and taking $V_i \to 0$ and $V_j \to 0 $, namely,
\begin{equation}
h_t({\mathbf x}_i,{\mathbf x}_j) = \lim_{V_i \to 0, V_j \to 0} \frac{\mathbb{E}_t[ n_{V_i} n_{V_j}|{\cal S}]}{V_i V_j}.
\label{def_correlations_h}
\end{equation}
The quantity $h_t({\mathbf x}_i,{\mathbf x}_j) d{\mathbf x}_i d {\mathbf x}_j$ is proportional to the probability of finding a pair of particles whose first member has coordinates ${\mathbf x}_i$ and the second has coordinates ${\mathbf x}_j$ at time $t$. Actually, it is customary to introduce the (dimensionless) normalized and centered pair correlation function $g$, which is obtained from $h$ by subtracting the product of the concentrations and the self-correlation and by dividing by the product of the concentrations~\cite{houchmandzadeh_pre_2009, natelson_theory}, namely,
\begin{equation}
g_t({\mathbf x}_i,{\mathbf x}_j) = \frac{h_t({\mathbf x}_i,{\mathbf x}_j) - \delta({\mathbf x}_i-{\mathbf x}_j) c_t({\mathbf x}_i)}{c_t({\mathbf x}_i) c_t({\mathbf x}_j) } -1.
\end{equation}

The evolution equations for these physical observables can be derived by resorting to a backward formalism. Calculations are rather cumbersome and are left to Appendix~\ref{backward_appendix}. The key result is that the physical observables can be formally obtained in terms of the Green's function ${\cal G}_t({\mathbf x}, {\mathbf x}_0)$ satisfying the backward equation
\begin{equation}
\frac{\partial}{\partial t} {\cal G}_t({\mathbf x}, {\mathbf x}_0) = {\cal L}^*_{{\mathbf x}_0}{\cal G}_t({\mathbf x}, {\mathbf x}_0),
\label{green_eq}
\end{equation}
with ${\cal G}_0({\mathbf x}, {\mathbf x}_0) = \delta({\mathbf x}- {\mathbf x}_0)$ and the boundary conditions of the problem at hand. We have defined the backward operator
\begin{equation}
{\cal L}^*_{{\mathbf x}_0}= D \nabla^2_{{\mathbf x}_0} + \lambda (\nu_1 - 1),
\label{backward_laplacian}
\end{equation}
where $\nu_1 = \sum_k k q_k$ is the average number of descendants per reproduction event.

By building upon the arguments discussed in Appendix~\ref{backward_appendix}, it can be shown that the concentration reads
\begin{equation}
c_t({\mathbf x}_i) = N \int_V d{\mathbf x}_0 p({\mathbf x}_0) {\cal G}_t({\mathbf x}_i, {\mathbf x}_0),
\label{eq_def_ave}
\end{equation}
which basically expresses a linear superposition of effects from single-particle contributions. As for the correlation function, from Appendix~\ref{backward_appendix} we get
\begin{align}
& h_t({\mathbf x}_i,{\mathbf x}_j) = \frac{N(N-1)}{N^2} c_t({\mathbf x}_i) c_t({\mathbf x}_j) + \delta({\mathbf x}_i-{\mathbf x}_j) c_t({\mathbf x}_i) \nonumber \\
& +\lambda \nu_2 \int_0^{t} dt' \int_V d{\mathbf x}' {\cal G}_{t'}({\mathbf x}_i, {\mathbf x}'){\cal G}_{t'}({\mathbf x}_j, {\mathbf x}') c_{t-t'}({\mathbf x}') ,
\end{align}
where $\nu_2 = \sum_k k(k-1) q_k $ is the second factorial moment of the number of descendants per reproduction event. For $N \gg 1$, $g_t({\mathbf x}_i,{\mathbf x}_j)$ reads then
\begin{align}
& g_t({\mathbf x}_i,{\mathbf x}_j) = \frac{\lambda \nu_2}{c_t({\mathbf x}_i) c_t({\mathbf x}_j)} \times \nonumber \\
& \int_0^{t} dt' \int_V d{\mathbf x}' {\cal G}_{t'}({\mathbf x}_i, {\mathbf x}'){\cal G}_{t'}({\mathbf x}_j, {\mathbf x}')c_{t-t'}({\mathbf x}').
\label{eq_def_corr}
\end{align}
Equation~\eqref{eq_def_corr} represents the (normalized) contributions to the correlations due to particles freely evolving from $t_0$ to $t-t'$, having a branching event in ${\mathbf x}'$ and whose descendants independently reach the points ${\mathbf x}_i$ and ${\mathbf x}_j$ at time $t$~\footnote{Though we have focused here on local observables, integral (i.e., spatially extended) estimators can be similarly defined so as to assess the relevance of the fluctuations: for a discussion, see Appendix~\ref{var_to_mean}.}. Observe that when either $\lambda=0$ or $q_0+q_1 = 1$ (i.e., $\nu_2 =0$) the normalized pair correlation function $g$ vanishes, which denotes the absence of correlations between particle positions in the absence of branching. Deviations of $g$ from zero are the signature of strong non-Poissonian fluctuations induced by the reproduction-disappearance mechanism~\cite{houchmandzadeh_pre_2009, young}. When $g \ge 1$, fluctuations are of the same order of magnitude as the concentration, which means that the information conveyed in the concentration alone (i.e., a pure deterministic approach) is useless. In particular, the presence of a peak at ${\mathbf x}_i \simeq {\mathbf x}_j$ in the shape of $g$, if any, would be the signature of particle clustering: because births can only occur at the location of a parent particle, walkers will preferentially gather together when the spatial smoothing due to diffusive mixing is weak. A peak in $g$ at short distances would thus mirror the enhanced probability of finding a pair of particles at very close spatial sites.

\section{Fluctuations around equilibrium}
\label{fluctuations_around_equilibrium}

Once the Green's function ${\cal G}_{t}({\mathbf x}, {\mathbf x}_0)$ of the problem has been determined, then by integration Eqs.~\eqref{eq_def_ave} and ~\eqref{eq_def_corr} allow explicitly characterizing the evolution of the particle concentration and of the correlations of the particle number, respectively. Formulas~\eqref{eq_def_ave} and ~\eqref{eq_def_corr} hold for any geometries and spatial source distributions~\footnote{Here for the sake of simplicity we have assumed that the initial number $N$ of particles is known; this hypothesis can be easily relaxed: see Appendix~\ref{source_def} for a discussion.}, and can accommodate arbitrary boundary and initial conditions (which affect the shape of ${\cal G}_{t}({\mathbf x}, {\mathbf x}_0)$).

A situation of particular interest is that of individuals prepared at equilibrium with respect to the spatial variable at time $t_0=0$. In this context, the case of unbounded domains has been discussed in detail by several authors~\cite{young, cox, houchmandzadeh_pre_2002, houchmandzadeh_pre_2009, houchmandzadeh_prl} and is briefly recalled in Appendix~\ref{scaling_limit} for the sake of completeness. Here we will focus on particles evolving in confined geometries, whose analysis can be carried out by resorting to the eigenfunction expansion of the Green's function. Under mild hypotheses  (see Appendix~\ref{modal_expansion} for a discussion), the Green's function of Eq.~\eqref{green_eq} can be represented in terms of a discrete sum of eigenfunctions $\varphi_{{\mathbf k}}$ of the operator ${\cal L}^*_{{\mathbf x}_0}$, in the form
\begin{equation}
{\cal G}_t({\mathbf x}, {\mathbf x}_0) = \sum_{{\mathbf k}} \varphi_{{\mathbf k}}({\mathbf x}) \varphi_{{\mathbf k}}({\mathbf x}_0) e^{\alpha_{\mathbf k} t},
\label{eigen_expansion}
\end{equation}
where $\alpha_{\mathbf k}$ are the associated eigenvalues~\cite{grebenkov}. The eigenvalues and the eigenfunctions depend on the specific boundary conditions. In most physical applications, one is often led to consider either (perfectly) reflecting or absorbing boundaries: in the former case, individuals reaching the walls bounce off and their trajectories are otherwise undisturbed (Neumann boundary condition); in the latter, individuals hitting the boundaries leak out and are thus lost (Dirichlet boundary condition). Neumann boundary condition would be representative, e.g., of neutrons multiplication in the presence of highly scattering shielding barriers, such as beryllium or heavy water~\cite{bell}, or the evolution of a bacterial colony confined on a Petri box with impermeable walls~\cite{houchmandzadeh_prl}. Absorbing boundaries are frequently met in radiation transport when the diffusing particles are free to escape upon crossing the external surface (the so-called geometrical leakage)~\cite{bell}.

Assuming that the individuals are prepared at equilibrium basically amounts to sampling the initial $N$-particle distribution on the fundamental spatial eigenstate of this system. In this case, we have $p({\mathbf x}_0) = p^{\text eq}({\mathbf x}_0) \propto \varphi_{{\mathbf 0}}({\mathbf x}_0)$, and we obtain
\begin{equation}
c^{\text eq}_t({\mathbf x}_i) = N p^{\text eq}({\mathbf x}_i) e^{\alpha_{\mathbf 0} t},
\label{eq_conc_equil_alpha}
\end{equation}
where $\alpha_{\mathbf 0}$ is the fundamental eigenvalue. Then, the spatial shape of the concentration would not vary, namely, $c_t({\mathbf x}_i) \sim p^{\text eq}({\mathbf x}_i)$, and its amplitude would evolve exponentially in time, with a rate $\alpha_{\mathbf 0}$. The sign of $\alpha_{\mathbf 0}$ determines the asymptotic behaviour of the concentration: when $\alpha_{\mathbf 0} > 0$ the population diverges in time and the system is said to be supercritical; when $\alpha_{\mathbf 0} <0$ the population shrinks to zero and the system is said to be subcritical. When $\alpha_{\mathbf 0} = 0$, the system is said to be critical and the concentration simplifies to $c^{\text eq}_t({\mathbf x}_i) = N p^{\text eq}({\mathbf x}_i)$, which means that, once prepared in the fundamental eigenstate, the system will stay in that eigenstate. Nuclear systems are typically operated around $\alpha_{\mathbf 0} = 0$, so as to have a constant power output~\cite{pazsit}.

This qualitative picture of the particle concentration for $\alpha_{\mathbf 0} = 0$ is in good agreement with the behaviour of the individuals displayed in Fig.~\ref{fig1} $a)$ and $b)$ for $q_1=1$, but seemingly not compatible with the behaviour observed in Fig.~\ref{fig1} $c)$ and $d)$, where for $q_0=q_2=1/2$ the Monte Carlo simulation shows that individuals are strongly clustered, and the density is far from being uniform. This contradiction is only apparent and stems from $c_t({\mathbf x}_i) d {\mathbf x}_i $ being formally defined as an ensemble average over an infinite number of realizations, whereas in real applications only a single realization is typically available and the particle concentration is usually defined as a spatial average over a region $V_i$~\cite{houchmandzadeh_pre_2009, young}. When the two kinds of expectations are equivalent, and a spatial sample is representative of the ensemble, the underlying stochastic process is said self-averaging: this assumption is satisfied for regular Brownian motion (with $\lambda = 0$, as in Fig.~\ref{fig1} $a)$ and $b)$), but is known to break down in the presence of branching~\cite{cox, young}. This shows the relevance of computing the higher moments of the particle number distribution. In order to go beyond the deterministic description and take into account fluctuations, the pair correlation function is needed. When the initial configuration is sampled on the fundamental eigenstate, from Eq.~\eqref{eq_def_corr} we get
\begin{align}
& g^{\text eq}_t({\mathbf x}_i,{\mathbf x}_j) =\frac{\lambda \nu_2 e^{- \alpha_{\mathbf 0} t}}{N p^{\text eq}({\mathbf x}_i)p^{\text eq}({\mathbf x}_j)} \times \nonumber \\
& \int_0^{t} dt' e^{-\alpha_{\mathbf 0} t'} \int_V d{\mathbf x}' p^{\text eq}({\mathbf x}') {\cal G}_{t'}({\mathbf x}_i, {\mathbf x}') {\cal G}_{t'}({\mathbf x}_j, {\mathbf x}').
\label{eq_corr_eigen_alpha}
\end{align}
Then, by using the eigenfunction expansion of the Green's functions and explicitly performing the time integral we are led to
\begin{align}
& g^{\text eq}_t({\mathbf x}_i,{\mathbf x}_j) =\frac{\lambda \nu_2 e^{- \alpha_{\mathbf 0} t}}{N p^{\text eq}({\mathbf x}_i)p^{\text eq}({\mathbf x}_j)} \times \nonumber \\
& \hspace{-1.4mm}\sum_{{\mathbf k}_i,{\mathbf k}_j} \frac{e^{(\alpha_{{\mathbf k}_i}+\alpha_{{\mathbf k}_j} - \alpha_{\mathbf 0})t}-1}{\alpha_{{\mathbf k}_i}+\alpha_{{\mathbf k}_j} - \alpha_{\mathbf 0}}A_{{\mathbf k}_i,{\mathbf k}_j}\varphi_{{\mathbf k}_i}({\mathbf x}_i)\varphi_{{\mathbf k}_j}({\mathbf x}_j),
\label{eq_corr_exp}
\end{align}
where the coefficients $A_{{\mathbf k}_i,{\mathbf k}_j}$ are given by
\begin{equation}
A_{{\mathbf k}_i,{\mathbf k}_j}=\int_V d{\mathbf x}' p^{\text eq}({\mathbf x}') \varphi_{{\mathbf k}_i}({\mathbf x}')\varphi_{{\mathbf k}_j}({\mathbf x}').
\label{coeff_A_k}
\end{equation}
Equation~\eqref{eq_corr_exp} has a fairly involved structure, though the case of Neumann boundary conditions leads to some simplifications (see Appendix~\ref{reflecting_boundaries}). In order to get some physical insight on the behaviour of the pair correlation function in bounded domains, it is convenient to perform a frequency analysis in the Laplace domain~\cite{williams}, namely,
\begin{equation}
g^{\text eq}_\omega({\mathbf x}_i,{\mathbf x}_j) = \int_0^\infty e^{-\omega t} g^{\text eq}_t({\mathbf x}_i,{\mathbf x}_j) dt.
\end{equation}
Without loss of generality, we can single out the fundamental mode, from which stems
\begin{align}
&g^{\text eq}_\omega({\mathbf x}_i,{\mathbf x}_j) =\frac{\lambda \nu_2 }{N p^{\text eq}({\mathbf x}_i)p^{\text eq}({\mathbf x}_j)} \frac{1}{\alpha_{\mathbf 0} + \omega}\frac{1}{\omega}\times \nonumber \\
& \Big[A_{{\mathbf 0},{\mathbf 0}}\varphi_{{\mathbf 0}}({\mathbf x}_i)\varphi_{{\mathbf 0}}({\mathbf x}_j)+\nonumber \\
& \hspace{-1.4mm}\omega \sum_{{\mathbf k}_i,{\mathbf k}_j \neq {\mathbf 0}} \frac{A_{{\mathbf k}_i,{\mathbf k}_j}}{2\alpha_{\mathbf 0}-\alpha_{{\mathbf k}_i}-\alpha_{{\mathbf k}_j} +\omega} \varphi_{{\mathbf k}_i}({\mathbf x}_i)\varphi_{{\mathbf k}_j}({\mathbf x}_j) \Big].
\label{omega_corr}
\end{align}
As expected on physical grounds, the overall intensity of the correlations is inversely proportional to the number of particles contained in the volume. The pre-factor $1/[(\alpha_{\mathbf 0}+\omega)\omega]$ determines the ultimate fate of the pair correlation function at long times (small $\omega$), and depends on the rate $\alpha_{\mathbf 0}$ at which the average population is increasing or decreasing. When the system is supercritical, i.e., $\alpha_{\mathbf 0} > 0 $, upon taking the inverse Laplace transform the pair correlation function for long times asymptotically converges to the constant
\begin{equation}
g^{\text eq}_{t \to \infty}({\mathbf x}_i,{\mathbf x}_j) \to \frac{\lambda \nu_2 }{N \alpha_{\mathbf 0}} {\cal M},
\end{equation}
where ${\cal M} = A_{{\mathbf 0},{\mathbf 0}}\varphi_{{\mathbf 0}}({\mathbf x}_i)\varphi_{{\mathbf 0}}({\mathbf x}_j) / p^{\text eq}({\mathbf x}_i)p^{\text eq}({\mathbf x}_j)$ is a normalization factor independent of the spatial coordinates. This means that fluctuations will be equally distributed at any spatial scale. In the supercritical regime, the average population is exponentially increasing at a rate $\alpha_{\mathbf 0}$, thus contributing to the mixing of the individuals: for sufficiently large $N$ one typically expects the amplitude of the pair correlation function to be $g \ll 1$, and fluctuations to be safely neglected. However, it may still happen that $g \ge 1$, when the number of initial particles is $N \ll \lambda \nu_2 {\cal M} / \alpha_{\mathbf 0}$. This can be understood as a competition between the growth rate $\alpha_{\mathbf 0}$ of the average population and the growth rate $\lambda \nu_2$ of branching-induced fluctuations: if $\alpha_{\mathbf 0}$ is rather small, strong correlations may have enough time to develop, despite the smoothing effect induced by the appearance of an increasing number of new diffusers. When the system is subcritical, i.e., $\alpha_{\mathbf 0} < 0 $, the pair correlation function at long times grows unbounded exponentially fast, as $g^{\text eq}_{t \to \infty}({\mathbf x}_i,{\mathbf x}_j) \sim \exp(-\alpha_{\mathbf 0} t)$: for negative $\alpha_{\mathbf 0} $, the average population is rapidly decreasing, which enhances the relative importance of fluctuations due to correlations. When the system is exactly critical, the pair correlation function asymptotically diverges with a linear scaling in time, namely,
\begin{equation}
g^{\text eq}_{t \to \infty}({\mathbf x}_i,{\mathbf x}_j) \sim \frac{\lambda \nu_2}{N} {\cal M} t.
\end{equation}
This linear scaling reflects the nature of the underlying Galton-Watson birth-death mechanism: when $\alpha_{\mathbf 0} =0$ a collection of $N$ individuals will go to extinction ($g \ge 1$) over a typical time $\sim N/\lambda$~\cite{houchmandzadeh_pre_2002, houchmandzadeh_pre_2009, zhang}.

The features displayed here are the signature of systems composed of a finite number of individuals in bounded geometries. The coefficients $A_{{\mathbf k}_i,{\mathbf k}_j}/(2\alpha_{\mathbf 0}-\alpha_{{\mathbf k}_i}-\alpha_{{\mathbf k}_j} +\omega)$ determine the relevance of the contributions of higher-order eigenfunctions to the spatial behaviour of the pair correlation function: the presence (or the absence) of a peak at short distances in $g^{\text eq}_t({\mathbf x}_i,{\mathbf x}_j)$ depends on these terms, which in turn depend on the eigenvalues of the system, hence on the geometry and on the physical parameters. Strictly speaking, according to the mathematical definition provided in~\cite{cox, houchmandzadeh_pre_2002, houchmandzadeh_pre_2009}, particle clustering may only occur for systems composed of an infinite number of individuals, and requires the short-distance peak of the pair correlation function to be divergent in time. In the present context, we actually use the term clustering in a loose sense, referring to the preferential appearance of fluctuations at short scales: in confined geometries, clustering (if any) is necessarily a transient regime, and we expect fluctuations to become spatially flat after the mixing time required by the particles to diffuse over the characteristic (finite) system size.

\section{One-dimensional box}
\label{one_dimensional}

In order to illustrate the formal approach proposed in the previous section, we consider here some simple examples of particle transport that yet retain the key features considered above. We assume that individuals are confined in a one-dimensional box of half-size $L$, i.e., a segment $[-L,L]$, with reflecting or absorbing boundary conditions imposed at the end points. Knowledge of the full eigenvalue spectrum allows determining the time scales that rule the fluctuations evolution.

\subsection{Reflecting boundaries}
\label{neumann}

Let us begin by the case of reflecting boundaries. Solving Eq.~\eqref{green_eq} on the one-dimensional box with Neumann boundary conditions $\partial_{x_0} {\cal G}_t(x, x_0) =0$ on $x_0 = \pm L$ yields the eigenfunction expansion
\begin{equation}
{\cal G}_t(x, x_0) = \frac{1}{2L}e^{\alpha_0 t} + \sum_{k=1}^\infty \varphi_k(x) \varphi_k(x_0) e^{\alpha_k t},
\end{equation}
where the eigenvalues $\alpha_k$ read
\begin{equation}
\alpha_k = - k^2\lambda_D  + \lambda (\nu_1-1)
\label{eigenvalues_ref}
\end{equation}
for $k \ge 0$, and we have introduced the quantity $\lambda_D = \pi^2 D/(2L)^2$, which is proportional to the mixing rate needed for the individuals to diffuse over the typical size of the box. The spatial eigenfunctions are
\begin{equation}
\varphi_k(x) = \frac{1}{\sqrt{L}} \cos\left(k \pi\frac{L-x}{2L} \right)
\end{equation}
for $k \ge 1$~\cite{carslaw, grebenkov}.

\begin{figure}[t]
\centerline{\epsfclipon \epsfxsize=9.0cm
\epsfbox{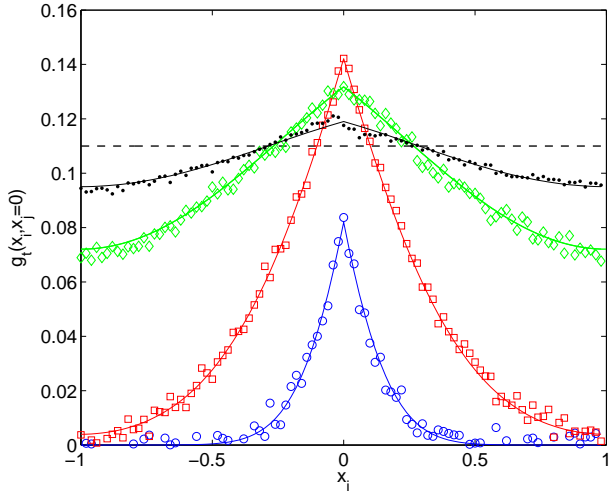} }
\caption{(Color online). The pair correlation function $g_t(x_i,x_j=0)$ for $N=100$ particles in a one-dimensional box of half-side $L$ with reflecting boundaries. The branching probabilities are $q_0=0.45$ and $q_2=0.55$, the diffusion coefficient is $D=0.01$, and the reproduction-disappearance rate is $\lambda = 1$. The mixing rate is then $\lambda_D \simeq 0.0162$, and the fundamental eigenvalue is $\alpha_0 = 0.1$ (supercritical regime). The pair correlation function is displayed at times $t=1$ (blue), $t=5$ (red), $t=20$ (green), and $t=30$ (black). Solid lines correspond to numerical integration, symbols to Monte Carlo simulations with $10^5$ realizations. The dashed line represents the asymptotic limit $\nu_2 / N (\nu_1-1) = 0.11$.}
\label{fig3}
\end{figure}

\begin{figure}[t]
\centerline{\epsfclipon \epsfxsize=9.0cm
\epsfbox{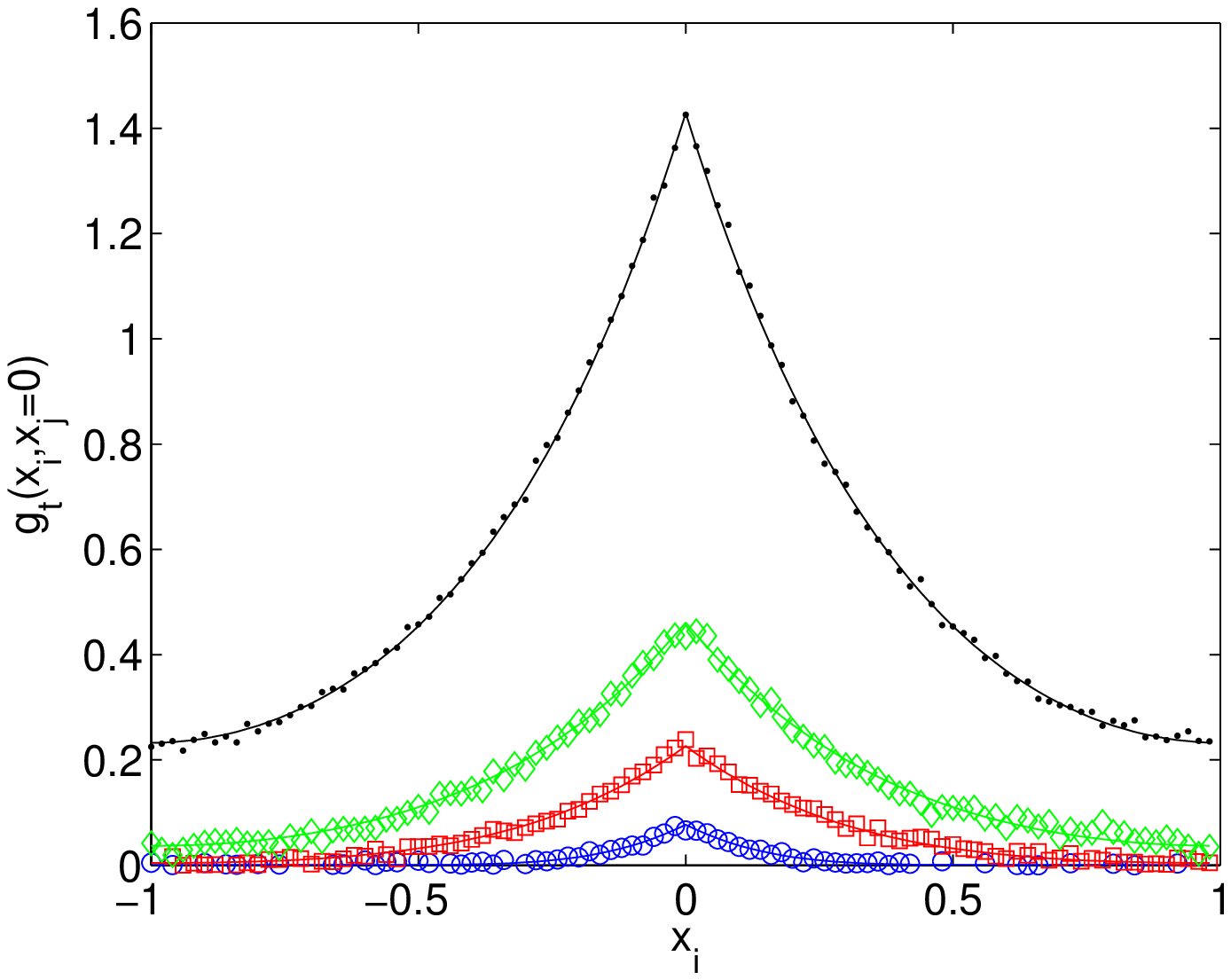} }
\caption{(Color online). The pair correlation function $g_t(x_i,x_j=0)$ for $N=100$ particles in a one-dimensional box of half-side $L$ with reflecting boundaries. The branching probabilities are $q_0=0.55$ and $q_2=0.45$, the diffusion coefficient is $D=0.01$, and the reproduction-disappearance rate is $\lambda = 1$. The mixing rate is then $\lambda_D \simeq 0.0162$, and the fundamental eigenvalue is $\alpha_0 = -0.1$ (subcritical regime). The pair correlation function is displayed at times $t=1$ (blue), $t=5$ (red), $t=10$ (green), and $t=20$ (black). Solid lines correspond to numerical integration, symbols to Monte Carlo simulations with $10^5$ realizations.}
\label{fig4}
\end{figure}

\begin{figure}[t]
\centerline{\epsfclipon \epsfxsize=9.0cm
\epsfbox{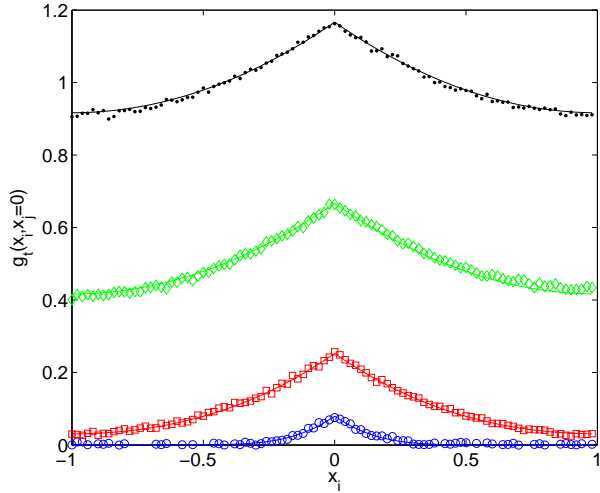} }
\caption{(Color online). The pair correlation function $g_t(x_i,x_j=0)$ for $N=100$ particles in a one-dimensional box of half-side $L$ with reflecting boundaries. The branching probabilities are $q_0=0.5$ and $q_2=0.5$, the diffusion coefficient is $D=0.01$, and the reproduction-disappearance rate is $\lambda = 1$. The mixing rate is then $\lambda_D \simeq 0.0162$, and the fundamental eigenvalue is $\alpha_0 = 0$ (critical regime). The pair correlation function is displayed at times $t=1$ (blue), $t=10$ (red), $t=50$ (green), and $t=100$ (black). Solid lines correspond to numerical integration, symbols to Monte Carlo simulations with $10^5$ realizations.}
\label{fig5}
\end{figure}

In order to assess the effects of fluctuations on the particle dynamics, let us assume that the $N$ branching Brownian particles are initially prepared at time $t_0 = 0$ at equilibrium~\footnote{For the sake of conciseness, in the following we drop the index `$eq$' and implicitly assume that all physical quantities refer to a system that is prepared on the fundamental eigenstate.}. Intuitively, this is achieved by assuming a uniform distribution, namely, $p(x_0) = 1/2L$, for $x_0 \in [-L,L] $. In this case, the concentration can be readily obtained from Eq.~\eqref{conc_eq_neumann}, and yields
\begin{equation}
c_t(x_i) = \frac{N}{2L} e^{\alpha_0 t},
\end{equation}
where $\alpha_0 = \lambda (\nu_1-1)$ is the fundamental eigenvalue. Then, the concentration is independent of $x_i$ and its amplitude grows or shrinks exponentially in time, at a rate $\alpha_0 $ depending only on $\lambda$ and $\nu_1$, regardless of the size of the box. When $\lambda = 0$, the concentration simply stays constant at $N/2L$. If $\lambda > 0$, the concentration grows unbounded ($\alpha_0>0$) for $\nu_1>1$, shrinks to zero ($\alpha_0<0$) for $\nu_1 < 1$ and is critical ($\alpha_0=0$) for $\nu_1 = 1$. As for the pair correlation function, from Eq.~\eqref{eq_equil_ref_corr} we get
\begin{align}
& g_t(x_i, x_j) =\frac{\lambda \nu_2}{N} \Big[ \frac{1-e^{-\alpha_0 t}}{\alpha_0} +\nonumber \\
&  2L e^{-\alpha_0 t} \sum_{k=1}^\infty \frac{e^{(2\alpha_k -\alpha_0)t }-1}{2\alpha_k -\alpha_0} \varphi_k(x_i) \varphi_k(x_j) \Big].
\label{corr_1d_ref}
\end{align}
Correlations trivially vanish when $\lambda = 0$ or $\nu_2 = 0$. When $\lambda >0$, by averaging Eq.~\eqref{corr_1d_ref} over the box, we obtain
\begin{equation}
\frac{1}{(2L)^2}\int_{-L}^L \int_{-L}^L dx_i dx_j g_t(x_i, x_j) =\frac{\lambda \nu_2}{N} \frac{1-e^{-\alpha_0 t}}{\alpha_0},
\label{corr_1d_ref_box}
\end{equation}
which means that the fluctuations affecting the total number of particles contained in the box (regardless of their positions) will saturate exponentially fast to a constant for positive $\alpha_0$, will diverge exponentially fast for negative $\alpha_0$, and will diverge linearly in time for an exactly critical system. The analysis of the spatial behaviour of Eq.~\eqref{corr_1d_ref} demands a closer inspection. By taking the Laplace transform of Eq.~\eqref{corr_1d_ref} and replacing the eigenvalues defined in Eq.~\eqref{eigenvalues_ref}, from Eq.~\eqref{omega_corr_ref} we get
\begin{align}
& g_\omega(x_i,x_j) =\frac{\lambda \nu_2 }{N } \frac{1}{\alpha_0 + \omega} \frac{1}{\omega}\times \nonumber \\
& \Big[1+\frac{\omega}{\lambda_D}\sum_{k = 1}^\infty \frac{2L}{2 k^2 +\frac{\omega}{\lambda_D}}\varphi_k(x_i) \varphi_k(x_j) \Big].
\label{omega_corr_ref_1d}
\end{align}
It is apparent that the ratio $\omega / \lambda_D$ is key to characterizing the space-dependent portion of $g_\omega(x_i,x_j)$. In particular, due to the competition between the birth-death rate $\alpha_0$ and the mixing rate $\lambda_D$, we expect the pair correlation function to display a rich behaviour. This is confirmed by numerical calculations: for illustration, we have computed $ g_t(x_i, x_j) $ for supercritical (Fig.~\ref{fig3}), subcritical (Fig.~\ref{fig4}) and critical (Fig.~\ref{fig5}) regimes and we have compared it to Monte Carlo simulations. In order to gain some physical insight, it is useful to single out distinct time scales. For $\omega \ll \lambda_D$, i.e., for times longer than the mixing time scale $1/\lambda_D$, the second term between square brackets in Eq.~\eqref{omega_corr_ref_1d} vanishes, and we have
\begin{align}
g_\omega(x_i,x_j) \sim \frac{\lambda \nu_2 }{N} \frac{1}{\alpha_0 + \omega} \frac{1}{\omega}.
\end{align}
Then, by taking the inverse Laplace transform, we recognize that the pair correlation function is spatially flat and asymptotically behaves as
\begin{align}
g_t(x_i,x_j) \sim \frac{\lambda \nu_2 }{N \alpha_0} \left(1- e^{-\alpha_0 t}\right).
\end{align}
This basically means that in this regime the system is behaving as a whole, and fluctuations affect any spatial scale. For positive $\alpha_0$, correlations at long times converge exponentially fast to a constant value, namely, $g_{t \to \infty}(x_i, x_j) \to \lambda \nu_2 /N \alpha_0 = \nu_2 / N (\nu_1-1)$, which is small for large $N$ (see Fig.~\ref{fig3}); yet, correlations may be relevant (i.e., $g_t(x_i, x_j) \ge 1$) whenever the initial number of particles is relatively small, namely, $N \le \nu_2/(\nu_1-1)$. For negative $\alpha_0$, correlations at long times grow unbounded exponentially fast (see Fig.~\ref{fig4}). Finally, for a critical system (Fig.~\ref{fig5}) correlations diverge linearly in time,
\begin{align}
g_t(x_i,x_j) \sim \frac{\lambda \nu_2 }{N} t.
\end{align}

In the limit $\omega \gg \lambda_D$, i.e., for times shorter than $1/\lambda_D$, the terms between square brackets in Eq.~\eqref{omega_corr_ref_1d} become important, and $g_\omega(x_i,x_j)$ has a non-trivial spatial shape. In this regime, the infinite sum in Eq.~\eqref{omega_corr_ref_1d} can be approximated by an integral by resorting to the Euler-Maclaurin formula, which leads to the closed-form expression
\begin{align}
& g_\omega(x_i,x_j) \sim \frac{\lambda \nu_2 }{N } \frac{1}{\alpha_0 + \omega}\frac{1}{\omega}\frac{\pi}{2}\sqrt{\frac{\omega}{2\lambda_D}}\times \nonumber \\
& \Big[e^{-\frac{\pi}{2}\sqrt{\frac{\omega}{2\lambda_D}} \frac{|x_i-x_j|}{L}}+e^{-\frac{\pi}{2}\sqrt{\frac{\omega}{2\lambda_D}} \frac{(2L-|x_i+x_j|)}{L}}\Big].
\label{omega_corr_ref_1d_nondiff}
\end{align}
For any given frequency $\omega$, $g^{\text eq}_\omega(x_i, x_j)$ displays a tent-like shape, symmetrical with respect to the line $x_i = x_j$. For fixed $x_i$, $g^{\text eq}_\omega(x_i,x_j)$ has a maximum at $x_j = x_i$. By virtue of the physical meaning of the pair correlation function, this behaviour reflects an enhanced probability of finding a pair of particles close to each other, which is the signature of clustering. Along the line $x_i=x_j$, $g^{\text eq}_\omega(x_i,x_j)$ is symmetrical with respect to $x_i=x_j=0$, where the function has a minimum, and the two global maxima are reached at the corners $x_j = x_i = \pm L$, which means that short-distance correlations are stronger when both particles are close to the boundaries of the box. Observe in particular that the short distance correlations for $x_i \simeq x_j \simeq \pm L$ (i.e., close to the boundaries) are twice as big as for $x_i \simeq x_j \simeq 0$ (i.e., at the center of the box).

Since $\omega \gg \lambda_D$, the exponential terms in Eq.~\eqref{omega_corr_ref_1d_nondiff} are rapidly decaying, so that we expect the relevant contributions to the correlations to come from particles being not too far apart, namely, $|x_i-x_j|/L \ll 1$. By choosing $x_i \simeq x_j \simeq 0$, which corresponds to short-distance correlations at the center of the box, $g_t(x_i,x_j)$ can be obtained by inverting the Laplace transform, and reads
\begin{align}
& g_t(x_i,x_j) \sim \frac{\lambda \nu_2 }{N } \frac{\pi}{2}\frac{1}{\sqrt{2 \alpha_0 \lambda_D}}e^{-\alpha_0 t} \text{erfi}(\sqrt{\alpha_0 t}),
\label{omega_corr_ref_1d_nondiff_time}
\end{align}
where $\text{erfi}(z)$ is the imaginary error function~\cite{erdelyi}. The critical case $\alpha_0=0$ yields
\begin{align}
& g_t(x_i,x_j) \sim \frac{\lambda \nu_2 }{N }\sqrt{\frac{\pi}{2} \frac{t}{\lambda_D}}.
\end{align}
In this regime, particles at the center of the box are not aware, yet, of the presence of the boundaries, so that we consistently recover the square root behaviour typical of one-dimensional critical systems in the thermodynamic limit (see Appendix~\ref{scaling_limit}). For $\alpha_0 \neq 0$, when $|\alpha_0| \ll \lambda_D$, i.e., when the growth rate due to reproduction and disappearance is much shorter than that of mixing, then the short-distance correlations at the center of the box at early time yield again
\begin{align}
& g_t(x_i,x_j) \sim \frac{\lambda \nu_2 }{N }\sqrt{\frac{\pi}{2} \frac{t}{\lambda_D}},
\end{align}
independent of $\alpha_0$, because particles in this regime are not sensitive to the fluctuations due to births and deaths. When on the contrary $|\alpha_0| \gg \lambda_D$, the effects due to reproduction and disappearance are in competition with mixing, and at longer times $1/|\alpha_0| \ll t \ll 1/\lambda_D$ the short-distance correlations at the center of the box yield
\begin{align}
& g_t(x_i,x_j) \sim \frac{\lambda \nu_2 }{N }\frac{\pi}{2}\Big[\frac{e^{-\alpha_0 t}}{\sqrt{2|\alpha_0| \lambda_D}} +\frac{1}{\alpha_0} \frac{1}{\sqrt{2\pi \lambda_D t}} \Big].
\end{align}

\subsection{Absorbing boundaries}
\label{dirichlet}

Solving Eq.~\eqref{green_eq} for the one-dimensional box with Dirichlet boundary conditions ${\cal G}_t(x, x_0) = 0$ on $x_0 = \pm L$ leads to an eigenfunction expansion for the Green's function in the form
\begin{equation}
{\cal G}_t(x, x_0) = \sum_{k=0}^\infty \varphi_k(x) \varphi_k(x_0) e^{\alpha_k t},
\end{equation}
where the eigenvalues $\alpha_k$ are
\begin{equation}
\alpha_k = - (k+1)^2 \lambda_D + \lambda (\nu_1-1),
\label{eigenvalues_abs}
\end{equation}
and the spatial eigenfunctions read
\begin{equation}
\varphi_k(x) = \frac{1}{\sqrt{L}} \sin\left((k+1)\pi\frac{L-x}{2L} \right),
\end{equation}
for $k \ge 0$~\cite{carslaw, grebenkov}. For this system, the analysis of the fluctuations due to a single particle born at $x_0$ at $t_0=0$ has been carefully carried out in~\cite{pazsit}. Here, similarly as done above for the reflecting boundaries, we would like to assess the impact of the particle number fluctuations on equilibrium, which for this system corresponds to the fundamental eigenstate $\sim \cos\left(\pi x/2L \right)$~\cite{grebenkov}. We sample then the initial $N$ particle positions on the normalized density
\begin{equation}
p(x_0) = \frac{\pi}{4L} \cos\left(\frac{\pi x_0}{2L} \right),
\end{equation}
for $x_0 \in [-L,L]$. Correspondingly, from Eq.~\eqref{eq_conc_equil_alpha} we can derive the concentration
\begin{equation}
c_t(x_i) = N \frac{\pi}{4L} \cos\left(\frac{\pi x_i}{2L} \right) e^{\alpha_0 t},
\end{equation}
where $\alpha_0 = \lambda (\nu_1-1) - \lambda_D$ is the fundamental eigenvalue. As expected, the spatial shape of the concentration does not evolve once the population is prepared on an equilibrium distribution, whereas the amplitude has a simple exponential behaviour with rate $\alpha_0$. Because of the spatial leakage, the sign of the fundamental eigenvalue $\alpha_0$ now depends on both the branching rate and on the mixing rate $\lambda_D$. When $\alpha_0 >0 $, the population growth due to branching is not sufficiently compensated by the spatial leakage from the boundaries, so that the concentration diverges exponentially (supercritical regime). This requires $\lambda (\nu_1 - 1) > \lambda_D$, with $\nu_1 >1$. On the contrary, when $\alpha_0 < 0 $ the concentration vanishes exponentially fast (subcritical regime) because either absorption dominates over reproduction, i.e., $\nu_1 < 1$, or because leakage dominates over branching, i.e., $1 < \nu_1 < 1 + \lambda_D/\lambda$. The critical regime is attained for $\alpha_0 = 0$, in which case the population stays constant: this is achieved whenever the physical parameters satisfy $\lambda (\nu_1-1) = \lambda_D$, with $\nu_1 > 1$, which precisely expresses the balance between branching and leakage. In particular, this implies the existence of a critical size $L_c$ of the box at which $\alpha_0 = 0$ for assigned values of the other physical parameters~\cite{pazsit}, namely,
\begin{equation}
L_c = \frac{\pi}{2} \sqrt{\frac{D}{\lambda (\nu_1-1)}}.
\end{equation}
As for the pair correlation function, from Eq.~\eqref{eq_corr_exp} we obtain
\begin{align}
& g_t(x_i, x_j) = \frac{\lambda \nu_2}{N} \Big[ \frac{32}{3\pi^2} \frac{1-e^{-\alpha_0 t}}{\alpha_0}+ \frac{e^{-\alpha_0 t}}{p(x_i) p(x_j)} \times \nonumber \\
& \sum_{k_i,k_j \neq 0} \frac{e^{(\alpha_{k_i}+\alpha_{k_j}-\alpha_0)t}-1}{\alpha_{k_i}+\alpha_{k_j}-\alpha_0} A_{k_i,k_j}  \varphi_{k_i}(x_i)\varphi_{k_j}(x_j) \Big],
\label{eq_corr_abs_1d}
\end{align}
where
\begin{align}
& A_{k_i,k_j} = \frac{1}{2L}\Big[\frac{\sin^2\left( (k_i+k_j-1)\frac{\pi}{2}\right)}{(k_i+k_j+1)(k_i+k_j+3)} \nonumber \\
& - \frac{\cos^2\left( (k_i-k_j)\frac{\pi}{2}\right)}{(k_i-k_j)^2-1} \Big]
\end{align}
from Eq.~\eqref{coeff_A_k}. We have moreover used
\begin{equation}
A_{0,0}\frac{\varphi_{0}(x_i)\varphi_{0}(x_j)}{p(x_i) p(x_j) }=\frac{32}{3\pi^2}.
\end{equation}
By taking the Laplace transform of Eq.~\eqref{eq_corr_abs_1d} and replacing the eigenvalues defined in Eq.~\eqref{eigenvalues_abs}, from Eq.~\eqref{omega_corr} we get
\begin{align}
& g_\omega(x_i,x_j) =\frac{\lambda \nu_2 }{N } \frac{32}{3\pi^2} \frac{1}{\alpha_0 + \omega} \frac{1}{\omega} \Big[1+ \frac{\frac{3}{2}L^2}{\cos(\frac{\pi x_i}{2L}) \cos(\frac{\pi x_j}{2L})} \times \nonumber \\
& \frac{\omega}{\lambda_D}\sum_{k_i,k_j\neq 0} \frac{A_{k_i,k_j}}{K_{k_i,k_j} +\frac{\omega}{\lambda_D}}\varphi_{k_i}(x_i) \varphi_{k_j}(x_j) \Big],
\label{eq_corr_abs_1d_laplace}
\end{align}
where we have used $K_{k_i,k_j}= k_i^2+k_j^2+2(k_i+k_j)$. In particular, for $\omega \ll \lambda_D$, i.e., for times longer than the mixing time scale, the space-dependent portion in Eq.~\eqref{eq_corr_abs_1d_laplace} becomes vanishing small, so that we expect the behaviour of systems with absorbing boundaries to be qualitatively similar to that of systems with reflecting boundaries (cf.~Eq.~\eqref{omega_corr_ref_1d}). In this regime, when $\alpha_0 > 0$ at long times the pair correlation function asymptotically converges to a constant value, namely, $ g_{t \to \infty}(x_i, x_j) \to (32/3\pi^2) \lambda \nu_2/\alpha_0 N $. When $\alpha_0 < 0$, correlations at long times grow unbounded exponentially fast. Finally, in the critical regime, with $\alpha_0 = 0$, the pair correlation function asymptotically diverges linearly in time.

When $\omega \gg \lambda_D$, i.e., for times shorter than the mixing time scale, numerical analysis shows that $g^{\text eq}_\omega(x_i, x_j)$ has again a tent-like shape, and for any given frequency $\omega$ displays a behaviour qualitatively similar to that observed for reflecting boundary conditions. In particular, short-distance correlations are stronger when both particles are close to the boundaries of the box.

\section{Conclusions}
\label{conclusions}

In this paper we have shown that the analysis of the fluctuations of a collection of Brownian particles subject to diffusion, reproduction and disappearance can be rather easily carried out by resorting to a backward formalism based on the Green's function. The proposed approach is fairly broad, and can accommodate arbitrary sources, geometries and boundary conditions. Special emphasis has been given to the case of initial conditions compatible with equilibrium. We have focused on the case of confined geometries with perfectly reflecting or absorbing boundaries: a generalization to the more involved case of mixed (Robin) boundary conditions, physically corresponding to partial absorption/reflection, would be straightforwardly achieved by correspondingly modifying the Green's function. We conclude by observing the proposed backward approach could be extended so as to include more general models of stochastic transport, such as L\'evy flights or Pearson random walks, provided that the underlying process is still Markovian: in this case, the major modification would consist in replacing the Laplacian with the appropriate backward transport operator associate to the process (for instance, the fractional Laplacian for L\'evy flights~\cite{laplacian} or the streaming operator for Pearson random walks~\cite{exp_flights}). Similarly, including space-varying parameters would be possible with minor changes.

\appendix

\section{The backward formalism}
\label{backward_appendix}

Consider a $d$-dimensional branching Brownian motion with diffusion coefficient $D$ and reproduction rate $\lambda$. A single walker starts from position ${\mathbf x}_0$ at time $t_0=0$. Let $n_{V_i}=n_{V_i}({\mathbf x}_0,t)$ be the number of particles that are found in a volume $V_i \subseteq V$ of the viable space when the process is observed at a time $t>t_0$. We are interested in determining the simultaneous detection probability ${\cal P}_t(n_{V_i}, n_{V_j}|{\mathbf x}_0)$ of finding $n_{V_i}$ particles in volume $V_i \subseteq V$ and $n_{V_j}$ particles in volume $V_j \subseteq V$, at time $t$, for a single particle starting at ${\mathbf x}_0$ at time $t_0$. It is convenient to introduce the associated two-volume probability generating function
\begin{equation}
W_t(u_i,u_j|{\mathbf x}_0) = \mathbb{E}[ u_i^{n_{V_i}({\mathbf x}_0,t)} u_j^{n_{V_j}({\mathbf x}_0,t)} ],
\end{equation}
from which the $m$-th (factorial) moments of $n_{V_i}$ and $n_{V_j}$ can be obtained by derivation with respect to $u_i$ and $u_j$, respectively. In particular, the average particle number reads
\begin{equation}
\mathbb{E}_t[ n_{V_i} | {\mathbf x}_0] = \frac{\partial}{\partial u_i} W_t( u_i,u_j| {\mathbf x}_0 ) \vert_{u_i=1,u_j=1}.
\end{equation}
For the two-volume correlations we take the mixed derivative, namely,
\begin{equation}
\mathbb{E}_t[ n_{V_i} n_{V_j}|{\mathbf x}_0] = \frac{\partial^2}{\partial u_i \partial u_j} W_t( u_i,u_j| {\mathbf x}_0 ) \vert_{u_i=1,u_j=1}.
\end{equation}
It can be shown~\footnote{See, for instance, the derivation in~\cite{pazsit, bell_nuc} or the sketch of proof in~\cite{zoia1}. In the context of reactor physics, the resulting equation is known as P\`al-Bell equation and is widely used for analysing the statistics of particle counting at a given detector~\cite{pazsit, bell_nuc}.} that $W_t(u_i,u_j|{\mathbf x}_0)$ satisfies the backward equation
\begin{equation}
\frac{\partial}{\partial t} W_t = D\nabla^2_{{\mathbf x}_0} W_t-\lambda W_t + \lambda G[ W_t ],
\label{backward_eq_W_one}
\end{equation}
where $G[z]$ is the probability generating function associated to the descendant number distribution, namely, $G[z]=\sum_k q_k z^k$. By taking the derivative of Eq.~\eqref{backward_eq_W_one} once we get the equation for the average particle number
\begin{equation}
\frac{\partial}{\partial t} \mathbb{E}_t[ n_{V_i}|{\mathbf x}_0]= {\cal L}^*_{{\mathbf x}_0}\mathbb{E}_t[ n_{V_i}|{\mathbf x}_0],
\label{eq_average_one}
\end{equation}
where the backward operator ${\cal L}^*_{{\mathbf x}_0}$ has been defined in Eq.~\eqref{backward_laplacian} and
\begin{equation}
\nu_1 =\frac{\partial}{\partial z} G[z]\vert_{z=1} = \sum_k k q_k.
\end{equation}
Equation~\eqref{eq_average_one} must be solved together with the initial condition $\mathbb{E}_0[ n_{V_i} |{\mathbf x}_0] = \mathbbm{1}_{V_i}({\mathbf x}_0).$

As for the correlations, by taking the mixed derivative of Eq.~\eqref{backward_eq_W_one} we obtain
\begin{align}
&\frac{\partial}{\partial t} \mathbb{E}_t[ n_{V_i} n_{V_j} |{\mathbf x}_0]=  \nonumber\\
&={\cal L}^*_{{\mathbf x}_0}\mathbb{E}_t[ n_{V_i} n_{V_j} |{\mathbf x}_0]+ \lambda \nu_2 \mathbb{E}_t[ n_{V_i}|{\mathbf x}_0] \mathbb{E}_t[n_{V_j} |{\mathbf x}_0] ,
\label{eq_correlations_one}
\end{align}
where
\begin{equation}
\nu_2 = \frac{\partial^2}{\partial z^2} G[z]\vert_{z=1} = \sum_k k(k-1) q_k.
\end{equation}
Equation~\eqref{eq_correlations_one} must be solved together with the initial condition $\mathbb{E}_0[ n_{V_i} n_{V_j} |{\mathbf x}_0] = \mathbbm{1}_{V_i}({\mathbf x}_0)\mathbbm{1}_{V_j}({\mathbf x}_0)$.

Equations~\eqref{eq_average_one} and~\eqref{eq_correlations_one} have both the general form
\begin{equation}
\frac{\partial}{\partial t} f_t({\mathbf x}_0)= {\cal L}^*_{{\mathbf x}_0}f_t({\mathbf x}_0) + a_t({\mathbf x}_0) ,
\end{equation}
where $a_t({\mathbf x}_0)$ is some known function with $a_0({\mathbf x}_0) = 0$ and $f_0({\mathbf x}_0) = b({\mathbf x}_0)$ for $t=0$, and admit the solution
\begin{align}
&f_t({\mathbf x}_0) =\int d{\mathbf x}' b({\mathbf x}') {\cal G}_t({\mathbf x}', {\mathbf x}_0)\nonumber\\
& + \int_0^{t} dt' \int d{\mathbf x}' a_{t'}({\mathbf x}') {\cal G}_{t-t'}({\mathbf x}', {\mathbf x}_0),
\end{align}
where ${\cal G}_t({\mathbf x}, {\mathbf x}_0)$ is the Green's function satisfying Eq.~\eqref{green_eq}. Then, for the average particle number we get
\begin{equation}
\mathbb{E}_t[ n_{V_i} |{\mathbf x}_0] =\int_{V_i} d{\mathbf x}'{\cal G}_t({\mathbf x}', {\mathbf x}_0).
\label{eq_ave_n}
\end{equation}
As for the correlations, we find
\begin{align}
\mathbb{E}_t[ n_{V_i} n_{V_j} |{\mathbf x}_0] = \int_{V_i \cap V_j} d{\mathbf x}' {\cal G}_t({\mathbf x}', {\mathbf x}_0) +\nonumber \\
\lambda \nu_2 \int_0^{t} dt' \int_V d{\mathbf x}' {\cal F}_{t'}(V_i,V_j,{\mathbf x}') {\cal G}_{t-t'}({\mathbf x}', {\mathbf x}_0),
\label{eq_corr_n}
\end{align}
where $V_i \cap V_j$ denotes the intersection of $V_i$ and $V_j$ and we have set
\begin{equation}
{\cal F}_{t}(V_i,V_j,{\mathbf x}) = \int_{V_i} d{\mathbf x}' {\cal G}_{t}({\mathbf x}', {\mathbf x}) \int_{V_j} d{\mathbf x}'' {\cal G}_{t}({\mathbf x}'', {\mathbf x}).
\end{equation}

Let us now consider a collection ${\cal S}$ of $N$ such individuals initially located at ${\mathbf x}^1_0$, ${\mathbf x}^2_0$, ${\mathbf x}^3_0$, $\cdots$, ${\mathbf x}^N_0$ at time $t_0=0$. Since particles evolve independently of each other, the contributions of each particle to the counting process $n_{V_i}= n_{V_i}({\mathbf x}^1_0, {\mathbf x}^2_0, {\mathbf x}^3_0, \cdots, {\mathbf x}^N_0,t)$ are additive, and the probability generating function satisfies
\begin{equation}
W_t(u_i,u_j | {\mathbf x}^1_0, {\mathbf x}^2_0, \cdots , {\mathbf x}^N_0) = \prod_{k=1}^N W_t(u_i, u_j| {\mathbf x}^k_0).
\end{equation}
Suppose that the initial positions are independently and identically distributed and obey the factorized density
\begin{equation}
P({\mathbf x}^1_0, {\mathbf x}^2_0, \cdots , {\mathbf x}^N_0) = \prod_{k=1}^N p({\mathbf x}^k_0) .
\end{equation}
The corresponding probability generating function $W_t(u_i,u_j | {\cal S})$ satisfies then
\begin{equation}
W_t(u_i,u_j | {\cal S}) = \prod_{k=1}^N \int_V d {\mathbf x}^k_0 p({\mathbf x}^k_0) W_t(u_i, u_j| {\mathbf x}^k_0),
\label{eq_source}
\end{equation}
which can be finally rewritten as
\begin{equation}
W_t(u_i, u_j | {\cal S}) = \left\langle W_t(u_i,u_j | {\mathbf x}_0) \right\rangle ^N,
\end{equation}
where we have denoted $\langle f({\mathbf x}_0) \rangle = \int_V d {\mathbf x}_0 p({\mathbf x}_0) f({\mathbf x}_0)$ the average over the distribution of the initial coordinates.

The moments of the $N$-particle observables can be again obtained as above. In particular, for the average particle number we get
\begin{equation}
\mathbb{E}_t[ n_{V_i}|{\cal S}] = N \left\langle \mathbb{E}_t[ n_{V_i} |{\mathbf x}_0] \right\rangle.
\label{N_ave}
\end{equation}
Hence, from Eqs.~\eqref{N_ave} and~\eqref{def_concentration_c} we obtain the local concentration
\begin{equation}
c_t({\mathbf x}_i) = N \int_V d{\mathbf x}_0 p({\mathbf x}_0) {\cal G}_t({\mathbf x}_i, {\mathbf x}_0) ,
\end{equation}
where we have used $\mathbbm{1}_{V_i}({\mathbf x})/V_i \to \delta({\mathbf x} - {\mathbf x}_i)$.

As for the correlations, we have
\begin{align}
& \mathbb{E}_t[ n_{V_i} n_{V_j}|{\cal S}] =N\left(N-1 \right)\left\langle \mathbb{E}_t[ n_{V_i} |{\mathbf x}_0] \right\rangle \left\langle \mathbb{E}_t[ n_{V_j} |{\mathbf x}_0] \right\rangle \nonumber \\
& + N \left\langle \mathbb{E}_t[ n_{V_i} n_{V_j} | {\mathbf x}_0] \right\rangle.
\label{N_corr}
\end{align}
Hence, from Eqs.~\eqref{N_corr} and~\eqref{def_correlations_h} we obtain
\begin{align}
& h_t({\mathbf x}_i,{\mathbf x}_j) = \frac{N(N-1)}{N^2} c_t({\mathbf x}_i) c_t({\mathbf x}_j) + \delta({\mathbf x}_i-{\mathbf x}_j) c_t({\mathbf x}_i) \nonumber \\
& +\lambda \nu_2 \int_0^{t} dt' \int_V d{\mathbf x}' {\cal G}_{t'}({\mathbf x}_i, {\mathbf x}'){\cal G}_{t'}({\mathbf x}_j, {\mathbf x}') c_{t-t'}({\mathbf x}') ,\nonumber
\end{align}
where we have used
\begin{equation}
\lim_{V_i \to 0, V_j \to 0} \frac{{\cal F}_{t}(V_i,V_j,{\mathbf x})}{V_i V_j} = {\cal G}_{t}({\mathbf x}_i, {\mathbf x}){\cal G}_{t}({\mathbf x}_j, {\mathbf x}).
\end{equation}

\section{Variance-to-mean ratio}
\label{var_to_mean}

A useful integral estimator so as to assess the entity of the fluctuations with respect to the average in a given region $V_i$ is the so-called variance-to-mean ratio $\chi$~\cite{cox_lewis, williams, pazsit}, which is defined as
\begin{equation}
\chi = \frac{\mathbb{E}_t[ n^2_{V_i} |{\cal S}] - \mathbb{E}_t[ n_{V_i} |{\cal S}]^2 }{\mathbb{E}_t[ n_{V_i}|{\cal S}]}.
\end{equation}
By replacing the definitions of $\mathbb{E}_t[ n^2_{V_i} |{\cal S}]$ and $\mathbb{E}_t[ n_{V_i} |{\cal S}]$, when $N \gg 1$ the variance-to-mean ratio can be expressed in terms of the Green's function, namely,
\begin{equation}
\chi = 1+\lambda \nu_2\frac{\int_0^{t} dt' \left\langle \int_V d{\mathbf x}' {\cal F}_{t'}(V_i,V_i,{\mathbf x}') {\cal G}_{t-t'}({\mathbf x}', {\mathbf x}_0) \right\rangle}{\left\langle \int_{V_i} d{\mathbf x}' {\cal G}_t({\mathbf x}', {\mathbf x}_0) \right\rangle}.\nonumber
\end{equation}
Observe that in the absence of reproduction-disappearance events ($\lambda = 0$), or for $q_0+q_1 = 1$ ($\nu_2=0$) the variance-to-mean ratio is identically equal to unit (i.e., fluctuations are Poissonian), which follows from the particle histories being uncorrelated. A departure from unit is the signature of non-Poissonian fluctuations due to correlations~\cite{houchmandzadeh_pre_2002, houchmandzadeh_pre_2009, houchmandzadeh_prl}. In the context of reactor physics, the variance-to-mean ratio is intimately related to the so-called Feynman alpha method~\cite{feynman}, which is used for the analysis of the correlations in neutron detectors due to fission chains~\cite{williams, pazsit}.

\section{Other kinds of sources}
\label{source_def}

In many practical applications, the initial number of particles is not known in advance and is itself a random quantity $M$, with distribution $Q(M)$. Assuming again independent and identically distributed coordinates ${\mathbf x}^k_0$, $k=1, 2, \cdots, M$, Eq.~\eqref{eq_source} can be then generalized by averaging over the realizations of $M$, namely,
\begin{align}
& W_t(u_i,u_j | {\cal S}_Q) =\nonumber \\ & \sum_M Q(M) \prod_{k=1}^M \int_V d {\mathbf x}^k_0 p({\mathbf x}^k_0) W_t(u_i, u_j| {\mathbf x}^k_0).
\label{source_poisson}
\end{align}
Often, the initial configuration is assumed to be a Poisson point process~\cite{pazsit, bell_nuc}, which means that the total number $M$ of starting particles obeys a Poisson distribution, i.e.,
\begin{equation}
Q(M)=\frac{\mu^M}{M!}e^{-\mu},
\end{equation}
where $\mu = \mathbb{E}[ M ]$ is the average number of source particles. In this case, using the independence property as above, the sum in Eq.~\eqref{source_poisson} can be explicitly carried out, which yields
\begin{equation}
W_t(u_i, u_j | {\cal S}_Q) = \exp \left( \mu \left\langle W_t(u_i,u_j | {\mathbf x}_0) -1 \right\rangle \right).
\end{equation}
This result takes the name of Campbell's theorem~\cite{bell_nuc}. In particular, if we choose $\mu = N$, by taking the derivatives of the probability generating function the average particle number would be left unchanged with respect to the case of fixed $N$ (as expected), namely, $\mathbb{E}_t[ n_{V_i} |{\cal S}_Q] = \mathbb{E}_t[ n_{V_i} |{\cal S}]$. As for the correlations, $\mathbb{E}_t[ n_{V_i} n_{V_j}|{\cal S}_Q]$ would be still given by Eq.~\eqref{N_corr}, provided that the factor $N(N-1)$ is replaced by $N^2$: this means that the correlations associated to a Poisson point source with $\mu=N$ would appreciably differ from those associated to a source with a fixed number $N$ of particles only when $N$ is relatively small.

\section{Thermodynamic limit}
\label{scaling_limit}

The so-called thermodynamic limit is attained by considering a large number $N$ of particles in a large volume $V$, and imposing that the ratio ${\cal C} = \lim_{N \to \infty, V \to \infty} N/V$ is finite. The Green's function for a $d$-dimensional infinite system is the Gaussian density
\begin{equation}
{\cal G}_t({\mathbf x}, {\mathbf x}_0) = \frac{e^{-\frac{|{\mathbf x}- {\mathbf x}_0|^2}{4 D t}+\lambda (\nu_1 -1) t}}{(4 \pi D t)^{d/2}},
\end{equation}
which spatially depends only on the relative particle distance $r=|{\mathbf x}- {\mathbf x}_0|$. As for the spatial density of the source particles, we take the uniform distribution $p({\mathbf x}_0) = 1/V$. From Eq.~\eqref{eq_def_ave}, the concentration then reads
\begin{equation}
c^\infty_{t}({\mathbf x}_i) = \lim_{N \to \infty, V \to \infty} c^{\text eq}_t({\mathbf x}_i) = {\cal C} e^{\lambda (\nu_1 -1) t},
\end{equation}
where we have used the normalization of the Gaussian density. At criticality, the concentration is stationary, namely, $c^\infty_{t}({\mathbf x}_i) = {\cal C}$. As for the pair correlation function, from Eq.~\eqref{eq_def_corr} we get
\begin{align}
& g^\infty_t(r) =\lim_{N \to \infty, V \to \infty} g^{\text eq}_t({\mathbf x}_i,{\mathbf x}_j) =\nonumber \\
& \frac{\lambda \nu_2 }{{\cal C} } e^{-\lambda (\nu_1 -1) t} \hspace{-1mm} \int_0^{t} \hspace{-1mm} dt'\frac{e^{-\frac{r^2}{8 D t'}+\lambda (\nu_1 -1) t'}}{(8 \pi D t')^{d/2}} ,
\end{align}
and we recover the result previously obtained in~\cite{houchmandzadeh_pre_2009}. In particular, for $\nu_1 = 1$ the integral in the pair correlation function can be carried out explicitly~\cite{houchmandzadeh_pre_2009}, and yields
\begin{equation}
g^\infty_t(r) =\frac{\lambda \nu_2 }{8 \pi^{d/2} D {\cal C}} r^{2-d} \Gamma\left(-1+ \frac{d}{2},\frac{r^2}{8Dt} \right),
\label{eq_inf_corr}
\end{equation}
where $\Gamma(a,z) = \int_z^\infty e^{-u} u^{a-1} du$ is the incomplete Gamma function~\cite{erdelyi}. The asymptotic time behaviour of Eq.~\eqref{eq_inf_corr} depends on the dimension $d$: it is known that $g^\infty_t(r) \sim \sqrt{t}$ for $d=1$, $g^\infty_t(r) \sim \log(t)$ for $d=2$, and $g^\infty_t(r) \sim \text{const}$ for $d>2$~\cite{houchmandzadeh_pre_2009}.

\section{Eigenfunction expansion}
\label{modal_expansion}

Generally speaking, when the domain $V$ is open, bounded and connected it is possible to solve for the Green's function of Eq.~\eqref{green_eq} by evoking the separation of variables~\cite{grebenkov}. If this is the case, then the Green's function ${\cal G}_t({\mathbf x}, {\mathbf x}_0)$ can be expanded in terms of a discrete sum of eigenfunctions $\varphi_{{\mathbf k}}$ of the operator ${\cal L}^*_{{\mathbf x}_0}$, in the form provided in Eq.~\eqref{eigen_expansion}~\cite{grebenkov}. We assume that such expansion is complete, which means that
\begin{equation}
\sum_{{\mathbf k}} \varphi_{{\mathbf k}}({\mathbf x}) \varphi_{{\mathbf k}}({\mathbf x}_0) = \delta({\mathbf x} - {\mathbf x}_0),
\end{equation}
and that the eigenvalues can be ordered so that $\alpha_{\mathbf 0} > \alpha_{\mathbf 1} \ge \cdots \ge \alpha_{\mathbf k} \ge \cdots$. In particular, if the fundamental eigenvalue is $\alpha_{\mathbf 0} = 0$ for a given choice of the physical parameters (depending also on the boundary conditions at $\partial V$), and the corresponding eigenstate is strictly positive, the system is said to be critical. The functions $\varphi_{{\mathbf k}}({\mathbf x})$ and $\varphi_{{\mathbf k}}({\mathbf x}_0)$ satisfy the boundary conditions and are ortho-normal, with
\begin{equation}
\int d{\mathbf x}' \varphi_{{{\mathbf k}_i}}({\mathbf x}') \varphi_{{{\mathbf k}_j}}({\mathbf x}') = \delta_{{{\mathbf k}_i},{{\mathbf k}_j}}.
\end{equation}

\section{Reflecting boundaries}
\label{reflecting_boundaries}

For reflecting (Neumann) boundary conditions the fundamental eigenstate being flat, and $\alpha_{\mathbf 0} = \lambda (\nu_1 -1)$~\cite{grebenkov}. Thus, if we choose $p^{\text eq}({\mathbf x}_0) = 1/V$, from Eq.~\eqref{eq_conc_equil_alpha} for the concentration we would simply have
\begin{equation}
c^{\text eq}_t({\mathbf x}_i) = \frac{N}{V} e^{\alpha_{\mathbf 0} t}.
\label{conc_eq_neumann}
\end{equation}
At criticality, $c^{\text eq}_t({\mathbf x}_i) = N/V$. As for the pair correlation function, from Eq.~\eqref{eq_corr_eigen_alpha} we obtain
\begin{equation}
g^{\text eq}_t({\mathbf x}_i,{\mathbf x}_j) =\lambda \nu_2 \frac{V}{N} e^{-\alpha_{\mathbf 0} t} \int_0^{t} dt' e^{-\alpha_{\mathbf 0} t'} {\cal G}_{2t'}({\mathbf x}_i, {\mathbf x}_j),
\label{corr_eq_neumann}
\end{equation}
where we have used the Markov property of the Green's functions, namely,
\begin{equation}
\int d{\mathbf x}' {\cal G}_{t}({\mathbf x}_i, {\mathbf x}'){\cal G}_{t}({\mathbf x}_j, {\mathbf x}') = {\cal G}_{2t}({\mathbf x}_i, {\mathbf x}_j).
\end{equation}
By resorting to the eigenfunction expansion, we get
\begin{align}
& g^{\text eq}_t({{\mathbf x}_i},{{\mathbf x}_j}) = \lambda \nu_2 \frac{V}{N} \times \nonumber \\
& e^{-\alpha_{\mathbf 0} t} \sum_{{\mathbf k} }\frac{e^{(2 \alpha_{\mathbf k}-\alpha_{\mathbf 0})t} - 1}{2 \alpha_{\mathbf k} - \alpha_{\mathbf 0}}\varphi_{{\mathbf k}}({\mathbf x}_i) \varphi_{{\mathbf k}}({\mathbf x}_j),
\label{eq_equil_ref_corr}
\end{align}
which we could have directly derived from Eq.~\eqref{eq_corr_exp} by imposing ortho-normality, i.e., $A_{{\mathbf k}_i,{\mathbf k}_j} = \delta_{{{\mathbf k}_i},{{\mathbf k}_j}}/V$. By singling out the fundamental mode and passing to the Laplace transform, from Eq.~\eqref{eq_equil_ref_corr} we have
\begin{align}
& g^{\text eq}_\omega({\mathbf x}_i,{\mathbf x}_j) =\frac{\lambda \nu_2 }{N } \frac{1}{\alpha_{\mathbf 0} + \omega}\frac{1}{\omega}\times \nonumber \\
& \hspace{-1.2mm} \Big[1+\omega \sum_{{\mathbf k} \neq {\mathbf 0}} \frac{V}{2\alpha_{\mathbf 0}-2\alpha_{{\mathbf k}} +\omega} \varphi_{{\mathbf k}}({\mathbf x}_i)\varphi_{{\mathbf k}}({\mathbf x}_j) \Big].
\label{omega_corr_ref}
\end{align}

\end{document}